\documentclass[preprint,showpacs,preprintnumbers,amsmath,amssymb,prb]{revtex4}

\usepackage{graphicx}
\usepackage{dcolumn}
\usepackage{bm}
\usepackage{units}
\usepackage{color}

\newcommand{\Ms}{M_{\mathrm{s}}}

\newcommand{\me}{m_{\mathrm{e}}}
\newcommand{\muB}{\mu_{\mathrm{B}}}

\newcommand{\Kx}{K_{\mathrm{x}}}
\newcommand{\Ky}{K_{\mathrm{y}}}

\newcommand{\TC}{T_{\mathrm{C}}}

\newcommand{\ux}{u_{\mathrm{x}}}

\newcommand{\je}{j_{\mathrm{e}}}
\newcommand{\Heff}{\vec{H}^{i}_{\mathrm{eff}}}

\newcommand{\bLL}{\beta_{\mathrm{LL}}}
\newcommand{\bG}{\beta_{\mathrm{G}}}
\newcommand{\uWG}{u_{\mathrm{Walker}}^{\mathrm{G}}}
\newcommand{\uWLL}{u_{\mathrm{Walker}}^{\mathrm{LL}}}

\begin{document}

\preprint{APS/123-QED}

\title{Current-induced domain wall motion including thermal effects based on Landau-Lifshitz-Bloch equation}

\author{C. Schieback}
\author{D. Hinzke}%
\author{M. Kl{\"a}ui}%
\author{U. Nowak}
 \email{ulrich.nowak@uni-konstanz.de}
\author{P. Nielaba}
\affiliation{%
Fachbereich Physik, Universit\"at Konstanz, Universit\"atsstra{\ss}e 10, 78457 Konstanz, Germany
}%

\date{\today}

\begin{abstract}
We employ the Landau-Lifshitz-Bloch (LLB) equation to investigate current-induced domain wall motion at finite temperatures by numerical micromagnetic simulations. We extend the LLB equation with spin torque terms that account for the effect of spin-polarized currents and we find that the velocities depend strongly on the interplay between adiabatic and non-adiabatic spin torque terms. As a function of temperature, we find non-monotonous behavior, which might be useful to determine the relative strengths of the spin torque terms experimentally.

\end{abstract}

\pacs{72.25.Ba Spin polarized transport in metals, 75.10.Hk Classical spin models, 75.60.Ch Domain walls and domain structure}
\maketitle

\section{\label{sec:level1} Introduction} 
Magnetic nanostructures in external magnetic fields as well as under the influence of spin-polarized currents have
become interesting research fields in recent years due to fundamental novel effects that occur for geometrically confined
spin structures, such as domain walls.\cite{klauiJP09}
Current-induced domain wall motion has been suggested as an alternative to the use of external magnetic fields
to induce switching, opening the possibility of simple device fabrication making field-generating strip lines redundant. While current-induced
domain wall motion is experimentally well established,\cite{yamaguchiPRL04,klauiPRL05} the underlying physical mechanisms are not completely understood yet and in particular the importance of the adiabatic and the non-adiabatic spin torque terms as well as domain wall transformations for high current densities are highly debated.\cite{heynePRL08,liPRL04,thiavilleEPL05} Furthermore, the influence of temperature on the effects has so far been neglected in the \unit[0]{K} calculations and so  the experimentally found temperature dependence of for instance the critical current densities is so far not understood.\cite{laufenbergPRL06}

To theoretically predict the behavior of a spin texture under current, one can numerically solve the Landau-Lifshitz-Gilbert (LLG) equation and computer simulations can be performed using either a micromagnetic model or a classical atomistic spin model. Spin torque effects can be taken into account
by including the adiabatic and the non-adiabatic torque terms.\cite{bergerJAP78,slonczewskiJMMM96,liPRL04,thiavilleEPL05} Due
to the computational expense of atomistic simulations, system sizes are restricted to a nanometer range, so that micromagnetic approaches are
desirable.  However, conventional micromagnetic calculations for larger system sizes lack the correct description of temperature effects because
of the assumption of a constant magnetization length. An alternative approach that has only recently started to be used widely to investigate realistic systems sizes
including temperature effects is to employ the so-called Landau-Lifshitz-Bloch equation.\cite{garaninPRB97} This equation forms the basis for
micromagnetic calculations at elevated temperatures using a macro-spin model where longitudinal relaxation processes are taken into
account \cite{kazantsevaPRB08} but so far the LLB equation has only been studied without the spin torque terms.

In this paper, we extend the LLB equation of motion by adding the spin torque terms and we study domain wall motion under the influence of current
and at variable temperatures. We determine the domain wall velocities and find that they exhibit a strong dependence on the temperature.
Furthermore, by the interplay between the adiabatic and the non-adiabatic spin torque the resulting onset of domain wall transformations (Walker breakdown)
is very sensitive to the temperature.

\section{Model} 
\subsection{Landau-Lifshitz-Bloch equation} 
\label{s:llb}
While in the LLG equation at \unit[0]{K} the length of the macro-spins stays constant, for finite temperatures an equation of motion for macro-spins allowing for
longitudinal relaxation was derived by Garanin \cite{garaninPRB97} within mean-field approximation from the
classical Fokker-Planck equation for atomistic spins interacting
with a heat bath.  The resulting ``Landau-Lifshitz-Bloch equation''
has been shown to be able to describe linear domain walls, a domain
wall type with non-constant magnetization length.\cite{koetzlerPRL93,kazantsevaPRL05,hinzkePRB08} Furthermore, the
predictions for the longitudinal and transverse relaxation times have been successfully compared with atomistic simulations
\cite{chubykaloPRB06} as well as rapid heating experiments.\cite{kazantsevaEPL09} Therefore, we now employ this equation to study the thermodynamics as well as the excitations of macro-spins due to currents.

The LLB equation can be written in the form
\begin{eqnarray}
   \dot{\vec{m}}_i= &-&
    \gamma \vec{m}_i \times
    \vec{H}^i_{\mathrm{eff}}-\frac{\gamma \alpha_{\perp}}{m_i^2}
    \vec{m}_i \times \Big(\vec{m}_i \times
    \vec{H}^i_{\mathrm{eff}}\Big) \nonumber \\ &+& \frac{\gamma
      \alpha_{\parallel}}{m_i^2}\Big( \vec{m}_i\cdot
      \vec{H}^i_{\mathrm{eff}}\Big)\vec{m}_i
\label{e:llb}
  \end{eqnarray}
where $\vec{m}_i$ is a to its zero temperature value normalized spin polarization and $\gamma$ the gyromagnetic ratio. The magnetization is not assumed to be of constant length and even its equilibrium value, $m_{\rm e}$, is temperature dependent. Hence, besides the usual precession and relaxation terms, the LLB equation contains another term which controls longitudinal relaxation.

The LLB equation is valid for finite temperatures and even above the
Curie temperature $\TC$ though the damping parameters and effective
fields are different below and above $\TC$. ${\alpha_{\parallel}}$ and
$\alpha_{\perp}$ are dimensionless longitudinal and transverse damping
parameters. For $T \leq \TC$ they are $ {\alpha_{\parallel}} = 2
\lambda T / (3 \TC)$ and $\alpha_{\perp} = \lambda (1 - T / (3
\TC))$. For $T \geq \TC$ the damping parameters are equal, $
\alpha_{\perp} = \alpha_{\parallel} = 2 \lambda T / (3 \TC)$. Here,
$\lambda$ is a microscopic damping parameter which characterizes the
coupling of the individual, atomistic spins to the heat bath. In the
limit $T \to 0$ the longitudinal damping parameter
${\alpha_{\parallel}}$ vanishes and with $\alpha_{\perp} = \lambda$
the LLB equation evolves into the usual Landau-Lifshitz (LL)
equation.

\begin{figure}
\includegraphics[width=8cm]{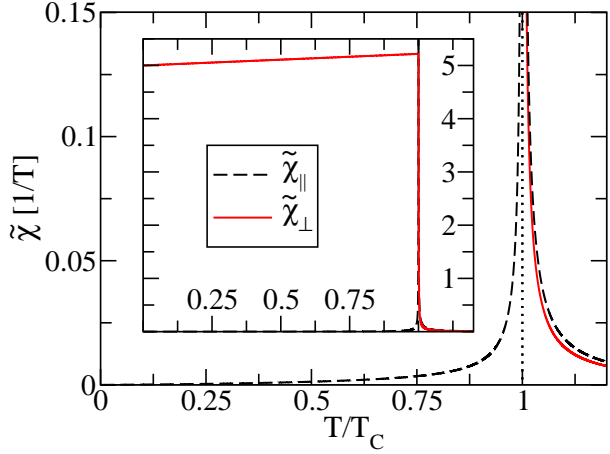}
\caption{\label{fig:susceptibility}(Color online) Equilibrium parallel and transverse susceptibility vs. temperature determined as explained in the text. The inset shows the transverse susceptibility vs. temperature on a larger scale.}
\end{figure}

\begin{figure}\includegraphics[width=8cm]{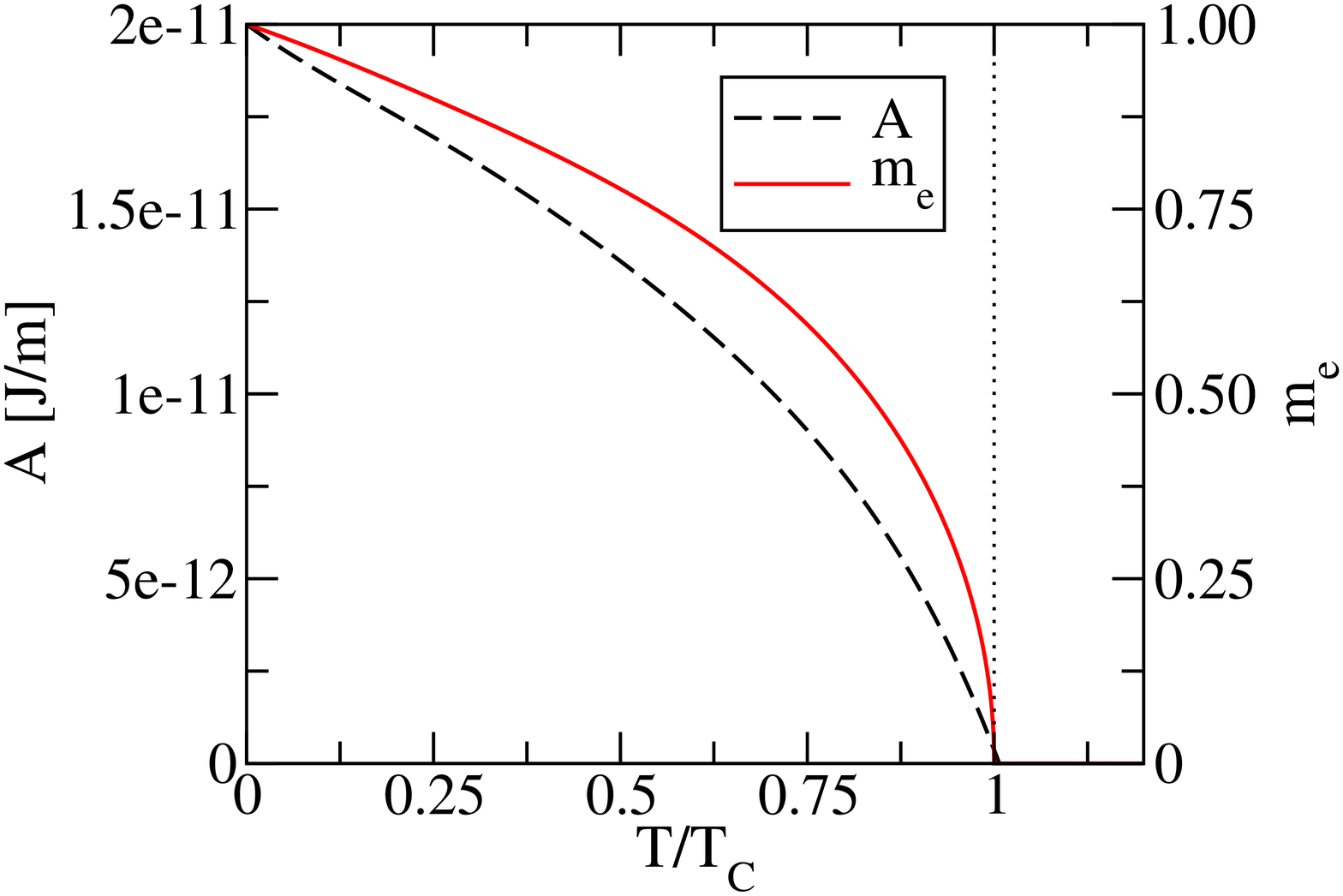}
\caption{\label{fig:exchange_me}(Color online) Exchange stiffness $A$ and reduced equilibrium magnetization $\me$ vs. temperature determined as explained in the text.}
\end{figure}

The effective fields of the LLB equation are the derivative  $\vec{H}^{i}_{\mathrm{eff}}=-\frac{1}{\Ms^0}\frac{\delta f}{\delta \vec{m}_i} $ of the free energy density $f$. The total field is given by \cite{garaninPRB97}
\begin{equation}
\vec{H}^{i}_{\mathrm{eff}}=  \vec{H}^{i}_{\mathrm{A}} + \vec{H}^{i}_{\mathrm{ex}}+ \left\{\!\!\! \begin{array}{lr}
  \frac{1}{2\tilde{\chi}_{\parallel}}\Big( 1 -
  \frac{m_i^2}{\me^2}\Big)\vec{m}_i & T \leq\TC
  \\ -\frac{1}{\tilde{\chi}_{\parallel}}\Big( 1 +
  \frac{2}{5}\frac{\TC m_i^2}{(T-\TC)} \Big)\vec{m}_i & T \geq
  \TC \end{array}\right.
    \end{equation}

with the biaxial anisotropy field,
\begin{equation}
\vec{H}^{i}_{\mathrm{A}} =  - \frac{1}{\tilde{\chi}_{\perp}}\left(\frac{1}{2}m^i_{y}\vec{e}_y +
    m^{i}_{z}\vec{e}_z\right),
\end{equation}
which makes the $x$-axis the easy axis, the $y$-axis the intermediate axis and the $z$-axis the hard axis of
the model. The exchange field is
\begin{equation}
\vec{H}^{i}_{\mathrm{ex}}  =   \frac{2 A}{\me^2 \Ms^0 \Lambda^2}
    \sum_j(\vec{m}_j - \vec{m}_i),
\end{equation}
were $\Lambda$ is the lateral size of the discretized cells, $A$ is the temperature dependent exchange stiffness, and $M_{\mathrm{s}}^0$ is the zero temperature saturation magnetization.  The susceptibilities $\tilde{\chi}_{l}$ are defined by $\tilde{\chi}_{l} = \partial m_l / \partial B_l$ with $l = \parallel, \perp$. Note, that at low temperatures the perpendicular susceptibility  $\tilde{\chi}_{\perp}$ is related to the temperature dependent anisotropy constant $K$ via  $\tilde{\chi}_{\perp} = M_{\mathrm s}^0 m_{\rm e}^2/(2K)$.\cite{garaninPRB97} We use functions for $\tilde{\chi}_{l}(T)$, $m_{\rm e}(T)$, and $A(T)$ as calculated before for the spin model (for details see  \cite{kazantsevaPRB08,hinzkePRB08}) but rescaled to reflect a ferromagnetic material with a Curie temperature of \unit[1043]{K} and  an $\Ms^0$ of $\unit[10^6]{A/m}$. Furthermore, we normalize the perpendicular susceptibility such that its value at \unit[0]{K}, $\tilde{\chi}_{\perp}(T=0)= \Ms^0/2\Kx$ corresponds to an anisotropy of $ \unit[\Kx=10^5]{J/m^3}$ and $2 \Ky=\Kx$. These functions are shown in Figs. \ref{fig:susceptibility} and \ref{fig:exchange_me}.

\subsection{Spin torque in the Landau-Lifshitz-Bloch equation} 
Throughout this paper we will consider a one-dimensional model of a domain wall. An established approach for the effect of a  spin-polarized current in the $x$-direction on a domain wall is presented in Ref.~\onlinecite{thiavilleEPL05}. In these studies the interaction between electron spins and magnetization has been treated by additional spin torque terms 

\begin{equation}
  \vec{T} =  - \ux  \frac{\partial \vec{S}}{\partial{x}} +
  \beta \vec{S} \times \ux \frac{\partial
    \vec{S}}{\partial{x}},
  \label{e:spin-torque}
\end{equation}
where $\vec{S}$ is a unit vector representing the direction of the magnetization. The first contribution to the spin torque is called the adiabatic term. It can be derived from an additional term in the magnetic free energy that takes into account the coupling of the magnetization to the spins of the electrical current,\cite{bazaliyPRB98} representing an adiabatic  transfer of angular momentum to the magnetization. In the adiabatic limit the spin polarization of the current is always oriented along the local direction of the magnetization. The second contribution is the non-adiabatic term that reflects the mistracking of the direction of the conduction electron spins with respect to the magnetization. It appears to play a role similar to the Gilbert damping term. Recent micromagnetic numerical investigations \cite{thiavilleEPL05,thiavilleJAP04,liPRL04,liPRB04,hePRB06} using a modified LLG equation including these terms have given a qualitative insight into the roles played by these two torque terms.

In the case of the current flowing in the $x$-direction, the magnitude of the effective spin current $\ux$ is given by $\ux =  j_{\mathrm{x}} /  M$ where $M$ is the magnetization and $j_{\mathrm{x}}$ the spin current density
$j_{\mathrm{x}} = \muB P \je / e$ which is proportional to the electrical current density $\je$ and to the polarization $P$. Here, $\muB$ is the Bohr magneton and $e$ the magnitude of the electron charge. Normally the spin current density is assumed to be temperature independent. In the following, we extend the model so that a temperature dependent spin current $j_{\mathrm{x}}(T)$  is taken into account. Under the assumption that the spin polarization is proportional to the magnetization, $P(M) = P^0 m$, and with $M = \Ms^0 m$, $\ux$ is $\ux =  \frac{P^0\je \muB}{\Ms^0 e}$ for all temperatures.

Under these assumptions the spin torque terms can be expressed in terms of the reduced magnetization $m$ as
\begin{equation}
  \vec{T} = - \ux \frac{\partial \vec{m}}{\partial{x}} +  \frac{\beta}{m} \vec{m} \times \ux \frac{\partial \vec{m}}{\partial{x}}.
  \label{e:spin-torque-LLB}
\end{equation}
Note, that the spin torque now is temperature dependent via the variable $m$ which, within the framework of the LLB equation, is no longer a unit vector of constant length. For the same reason, the adiabatic torque term can no longer be expressed as a double cross product. Instead it is
\begin{equation}
	- \ux \frac{\partial \vec{m}}{\partial{x}}  = \frac{\ux}{m^2} \Big( \vec{m} \times (\vec{m} \times \frac{\partial \vec{m}}{\partial{x}})
	-  \vec{m} \cdot(\vec{m} \cdot \frac{\partial \vec{m}}{\partial{x}}) \Big),
\end{equation}
which means that the adiabatic term, within the LLB equation, gives rise to an additional longitudinal spin torque term which vanishes in the LLG equation due to the assumption of a constant length of the magnetization vector. As pointed out in Ref.~\onlinecite{haneyARXIV09} this term corresponds to the spin accumulation and in the metal systems considered here, it constitutes usually a small effect.

It is not yet fully clear whether the spin torque term $\vec{T}$ should be added to the Landau-Lifshitz or the Landau-Lifshitz-Gilbert form of the equation of motion (for more details see the discussion in Refs.~\onlinecite{stilesPRB07,smithPRB08,stilesPRB08}). The same problem arises with the LLB equation. In the following, we extend the LLB equation with both forms of the damping, the one after Landau and Lifshitz as well as the one after Gilbert. The LL form of the LLB equation (Eq.~\ref{e:llb}) is the original one as derived by Garanin.\cite{garaninPRB97} Eq.~\ref{e:llb} now reads with the additional spin torque terms from Eq.~\ref{e:spin-torque-LLB} 
\begin{eqnarray}
   \dot{\vec{m}}_i=&-&  \gamma \vec{m}_i \times \Heff - \frac{\gamma\alpha_{\mathrm{\perp}}}{m^2_i} \vec{m}_i \times \Big(\vec{m}_i \times \Heff\Big)\nonumber \\ &+& \frac{\gamma\alpha_{\mathrm{\parallel}}}{m^2_i}\Big(\vec{m}_i\cdot  \Heff\Big)\vec{m}_i \nonumber \\
&-&\ux \frac{\partial \vec{m}_i}{\partial x} +  \frac{\bLL}{m_i} \vec{m}_i \times \ux \frac{\partial \vec{m}_i}{\partial x}.
\label{e:LL-LLB}
\end{eqnarray}
Neglecting terms of the order of $\alpha^2$ the LLB equation can be transformed into an equation with a damping term following Gilbert. Adding the same spin torque terms $\vec{T}$ to this form of the LLB equation yields
\begin{eqnarray}
        \dot{\vec{m}}_i = &-& \gamma \vec{m}_i \times\Heff + \frac{\gamma  \alpha_{\mathrm{\perp}}}{m^2_i} \vec{m}_i \times  \dot{\vec{m}}_i \nonumber \\
& + & \frac{\gamma\alpha_{\mathrm{\parallel}}}{m^2_i} \Big( \vec{m}_i\cdot \Heff\Big)\vec{m}_i \nonumber \\
 &-& \ux \frac{\partial \vec{m}_i}{\partial x} +  \frac{\bG}{m_i} \vec{m}_i \times \ux \frac{\partial \vec{m}_i}{\partial x}.
\label{e:LLG-LLB}
\end{eqnarray}
Note, that we use the notation $\bG$, $\bLL$ for the non-adiabatic prefactor only for convenience in order to distinguish in the following between the LL and LLG form of the LLB equation.

In the next step we transform Eq.~\ref{e:LLG-LLB} into an explicit form so that we are able to compare it with Eq.~\ref{e:LL-LLB} and also since an explicit equation is more convenient for a numerical treatment. This explicit equation can be derived once again neglecting terms of the order $\alpha^2$ and $\alpha \beta$ and it is given by
\begin{eqnarray}
  \dot{\vec{m}}_i = &-& \gamma \vec{m}_i \times \Heff - \frac{\gamma \alpha_{\mathrm{\perp}}}{m^2_i} \vec{m}_i \times \Big(\vec{m}_i \times \Heff\Big) \nonumber \\
&+& \frac{\gamma\alpha_{\mathrm{\parallel}}}{m^2_i}\Big(\vec{m}_i\cdot  \Heff \Big)\vec{m}_i \nonumber \\
&-&\ux \frac{\partial \vec{m}_i}{\partial x} +  \Big(\frac{\bG}{m_i} -\frac{\alpha_{\mathrm{\perp}}}{m^2_i}\Big)\vec{m}_i \times \ux \frac{\partial \vec{m}_i}{\partial x}.
\label{e:ex-LLG-LLB}
\end{eqnarray}
The only difference between Eq.~\ref{e:LL-LLB} assuming Landau-Lifshitz damping, and Eq.~\ref{e:ex-LLG-LLB} assuming Gilbert damping, is the prefactor of the last term. Eqs.~\ref{e:LL-LLB} and \ref{e:ex-LLG-LLB}  are mathematically identical for $\bLL = \bG - \alpha_\perp/m_i$.
Note, that at \unit[0]{K}  both equations evolve into the well established explicit versions of the LL respectively LLG equation with spin torque terms.\cite{stilesPRB07,smithPRB08,stilesPRB08}

\section{Analytical Model}
In the following, we present analytical calculations for a one-dimensional system aligned along the $x$-direction. We extended well-established, analytical \unit[0]{K}  calculations \cite{thiavilleJAP04,thiavilleEPL05} to elevated temperatures using the LLB equation as equation of motion. Note, that within the LLB approach domain wall profiles change with increasing temperature first from circular to elliptical and later on to linear.\cite{koetzlerPRL93,kazantsevaPRL05,hinzkePRB08} However, these effects occur only at higher temperatures close to $\TC$ (depending on the strength of the anisotropy), and are neglected in our analytical calculations. Hence, we assume a temperature independent domain wall type, which here is a transverse domain wall along the $x$-direction, which does not change its spin structure dynamically. This assumption, is later tested by comparison with numerical calculations without these approximations.

The assumed domain wall profile is as usual\cite{kazantsevaPRL05,hinzkePRB08} 
\begin{equation}
 \vec{m} = -\me \Big\{\tanh\Big(\frac{x}{\Delta}\Big)\vec{e}_x+\frac{\cos(\phi)}{\cosh\big(\frac{x}{\Delta}\big)}\vec{e}_y +
\frac{\sin(\phi)}{\cosh\big(\frac{x}{\Delta}\big)}\vec{e}_z\Big\}
\label{e:DWprofile}
\end{equation}
with the temperature dependent domain wall width
\begin{equation}
\Delta(T, \phi) = \frac{2}{\Ms^0}\sqrt{\frac{A(T)\tilde{\chi}_{\perp}(T)}{\me^2(T)} \frac{1}{(1+\sin^2\phi)}}
\label{e:DWwidth}
\end{equation}
 and the out-of-plane angle $\phi$. In the low temperature limit  this equation has the well known form,
 $\Delta = \sqrt{2A/(K(1+\sin^2{\phi}))}$.

The equations of motion are calculated as described in Ref.~\onlinecite{slonczewskiJMMM96}.  We assume Gilbert damping first, simply because it is the more common assumption in connection with spin torque calculations. For the domain wall profile above  the equations of motion for the position of the domain wall $x_\mathrm{G}$ and the angle $\phi$ are:
\begin{eqnarray}
  \dot{x_{\mathrm{G}}} &=& \frac{\me \Delta \gamma}{4\tilde{\chi}_{\perp}}\sin2\phi + \ux \\
  \dot{\phi} &=& -\frac{\alpha_{\perp}\gamma}{4\tilde{\chi}_{\perp}}\sin2\phi - \frac{(\alpha_{\perp} - \bG \me)} {\me}\frac{\ux}{\Delta}
\end{eqnarray}
These equations are calculated based on the original procedure (see \cite{thiavilleEPL05} and references therein).
The corresponding Walker threshold $\uWLL$~\cite{schryerJAP74} can be calculated under the assumptions that  $\dot{\phi} = 0$ and that the domain wall width  $\Delta({\phi,T})$ reaches its minimum $\Delta_{\mathrm{min}}$ at $\phi = \pm \pi/4$. The Walker threshold is then given by
 \begin{equation}
   \uWG=\frac{\gamma}{4 \tilde{\chi}_{\perp}} \Delta_{\mathrm{min}} \frac{\alpha_{\perp} \me}{|\alpha_{\perp}-\bG \me|} 
    \label{e:uWalker_G}
 \end{equation}
and the average domain wall velocity is 
\begin{equation}
   \langle v \rangle_{\mathrm{G}} = \frac{\bG\ux\me}{\alpha_{\perp}} \pm  \frac{\me \Delta_{\mathrm{min}} \gamma}{4\tilde{\chi}_{\perp}} \sqrt{\big(\frac{\ux}{\uWG}\big)^2 -1 },
        \label{e:avgv_G}
\end{equation}
with the plus sign for $(\alpha_{\perp} - \bG \me) \ux > 0$ and the minus sign otherwise. This equation contains a contribution which is linear in the current (in the non-adiabatic case) and a second square-root contribution above the Walker threshold.   The temperature dependence is included in the temperature dependent parameters $A(T), \me(T), \chi_{\perp}(T)$, and $\alpha_{\perp}(T)$. At zero temperature these results are identical to those gained with the LLG equation.\cite{thiavilleJAP04,thiavilleEPL05} The longitudinal spin torque does not affect the analytical results since we assume a constant domain wall type. Very close to the Curie temperature this assumption is no longer valid and deviations can be expected, which are beyond the scope of the current investigation.

\begin{figure}
\includegraphics[width=8cm]{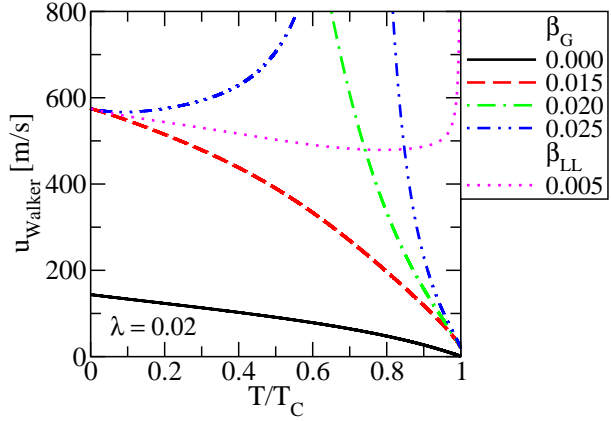}
\caption{(Color online) Walker threshold $u_{\mathrm{Walker}}$ according to Eq.~\ref{e:uWalker_G} vs. the reduced temperature $T/T_{\rm C}$ for different values of the non-adiabatic prefactor $\bG$ as well as $\bLL$.}
\label{fig:fig_walker_breakdown}
\end{figure}

In Fig.~\ref{fig:fig_walker_breakdown} the Walker threshold according to Eq.~\ref{e:uWalker_G} is shown as a function of the reduced temperature  ($T/\TC$). In all cases the Walker threshold vanishes at the Curie temperature. In the adiabatic case ($\bG=0$) and in general for $\bG < \lambda = \alpha_{\perp}(T=0)$ it decreases with increasing temperatures. For $\bG = \lambda = 0.02$ the Walker threshold diverges approaching zero temperature. The behavior for $\bG > \lambda$ is similar, but here the Walker threshold diverges at a finite temperature. As we will see in the following, the understanding of the temperature dependence of the Walker threshold is the key for understanding the temperature dependence of the domain wall velocity.

The equation of motion for Landau-Lifshitz damping (Eq.~\ref{e:ex-LLG-LLB}) can be calculated and solved in the same way as for the Gilbert damping with corresponding results for the Walker threshold and the average domain wall velocity for $\bLL = \bG - \alpha_\perp/\me$. Note, however, that since this transformation is temperature dependent, the temperature dependence of both, Walker threshold and domain wall velocity are different for Landau-Lifshitz and Gilbert damping, respectively.  The temperature dependence of the Walker threshold assuming Landau-Lifshitz damping is also shown in Fig.~\ref{fig:fig_walker_breakdown}. The main difference is that approaching the Curie temperature the Walker threshold does not vanish but diverges. In the limit of low temperatures, however, the Walker threshold converges to the one following Gilbert damping.

\section{Numerical Model} 
By means of computer simulations, temperature dependent domain wall velocities were calculated for a one-dimensional system of \unit[512]{nm} length, discretized with one nm cell size. The initial magnetization configuration was a planar domain wall positioned in the middle of the chain with the temperature dependent profile and width given by Eqs.~\ref{e:DWprofile} and \ref{e:DWwidth} with $\phi=0$. For \unit[0]{K} the domain wall width varies between $\Delta_{\mathrm{max}} = \unit[20]{nm}$ ($\phi=0$) and $\Delta_{\mathrm{min}} = \unit[14.26]{nm}$ ($\phi=\pi/4$). At the ends of the system the spins were fixed as boundary conditions in the $x$-direction and -$x$-direction, respectively. To minimize the influence of these boundary condition the domain wall was only allowed to move within a range of \unit[60]{nm} from the center of the system. When the domain wall moved out of this interval, it was shifted back along the $x$-coordinate and re-positioned at the opposite side of the interval. The domain wall velocity was calculated from the derivative of the spatially averaged $x$-component of the magnetization versus time. The numerical time integration of Eqs.~\ref{e:LL-LLB} and \ref{e:ex-LLG-LLB} was carried out using a Heun-method~\cite{garciaPRB98,nowakARCP01} with \unit[1.8]{fs} time step size.

\section{Results}               
\subsection{Adiabatic spin torque effect}             

\begin{figure}
\includegraphics[width=8cm]{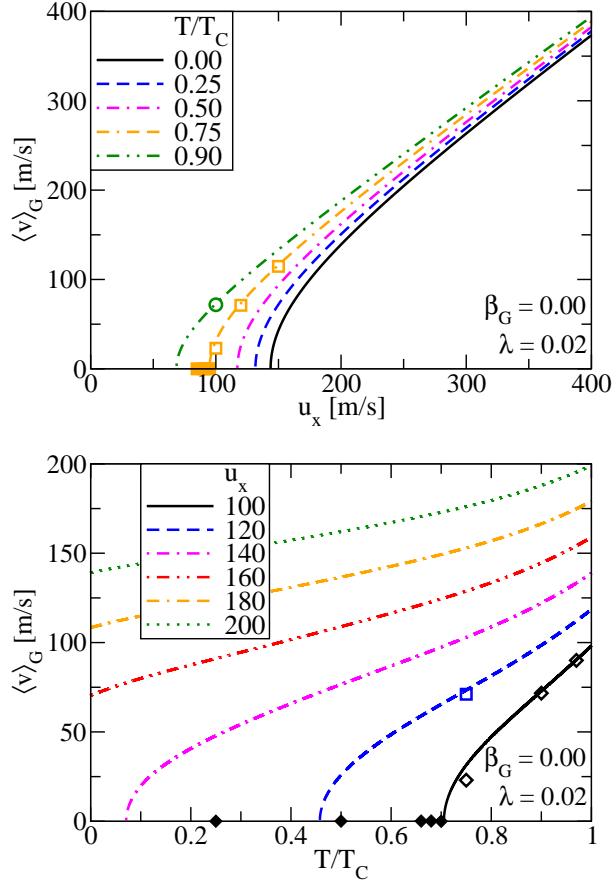}
\caption{(Color online) The top figure shows the average domain wall velocity $\langle v \rangle_{\rm G}$ calculated from the LLB equation
  with Gilbert damping vs. the effective spin current $u_x$ for different temperatures. The solid line for $T=\unit[0]{K}$ represents the case where
  the LLB and the LLG equation are identical. The bottom figure shows $\langle v \rangle_{\rm G}$ vs. the reduced temperature $T/T_{\rm C}$. The lines represent the
  analytic solution (Eq.~\ref{e:avgv_G}) and the points are from numerical simulations. The calculations are for the adiabatic case, $\beta_{\rm G} = 0$ and $\lambda = 0.02$.}
\label{fig:dwVelocity_bG--0.000.eps}
\end{figure}

In the following, first the pure adiabatic spin torque effect ($\bG = 0$) is discussed. For zero temperature, this effect was already investigated previously based on the LLG equation.\cite{thiavilleJAP04,tataraPRL04,liPRB04,schiebackEPJB07} It was found that for low effective spin currents $\ux$ the domain wall moves along the wire until it reaches a maximum displacement where it stops. At the same time, the magnetization of the domain wall is tilted out of the easy plane up to a maximum out-of-plane angle. This behavior can be explained by an analysis of the different terms of the extended LLG equation (Eq.~\ref{e:ex-LLG-LLB}): the first spin torque term which moves the domain wall is balanced by an "internal" torque due to the anisotropy contribution to the effective field. The displacement of the domain wall in $x$-direction is eventually stopped by the precessional term acting in the opposite direction, while the second spin torque term which is responsible for tilting the magnetization out of the easy plane is balanced by the damping term.

It was even analytically predicted \cite{thiavilleJAP04,tataraPRL04,liPRB04} that the averaged domain wall velocity as a function of the effective spin current remains zero unless the current exceeds a critical value $u_{\rm c}$. This predicted critical current was also found in atomistic simulations  at \unit[0]{K}.\cite{schiebackEPJB07} Below the critical effective spin current  ($u_x < u_{\rm c}$) no continuous domain wall motion is observed while above the critical current the spin torque term can no longer be balanced by the anisotropy. Consequently domain wall motion occurs in addition to a precession of the magnetization around the $x$-axis.

We find the same behavior in the extended LLB equation. This can be seen in Fig. \ref{fig:dwVelocity_bG--0.000.eps} where the averaged domain wall velocity is shown as a function of the effective spin current $u_x$ for different reduced temperatures $T/\TC$.  However, it is found that the critical effective spin current is temperature dependent, following the equation
 \begin{equation}
   u_{\rm c} = \frac{\gamma \me}{4 \tilde{\chi}_{\perp}} \Delta_{\mathrm{min}}
    \label{e:uWalker_G_0}
 \end{equation}
(Eq.~\ref{e:uWalker_G} for $\bG=0$). For larger temperatures the critical current decreases since thermodynamically the anisotropy decreases.  This leads to the fact that domain walls at higher temperatures are faster than at low temperatures.

Furthermore, in the bottom part of Fig. \ref{fig:dwVelocity_bG--0.000.eps} the averaged domain wall velocity is shown as a function of the reduced temperature for different values of the effective spin current.  In the limit of low current a critical temperature $T^{\star}$ can be identified.  For $T < T^{\star}$ no continuous domain wall motion is observed while for $T > T^{\star}$ domain wall motion occurs. This critical temperature is shifted to lower values for higher spin currents. In the limit of high effective spin current $T^{\star}$ vanishes and  domain wall motion can be observed over the whole temperature range. Here, the spin torque effect is no longer balanced by the anisotropy and only the terms responsible for the precession of the magnetization around the $x$-axis affects the domain wall motion.

In both figures, analytical curves and numerical results agree. This demonstrates clearly that the assumption made for the derivation of Eq.~\ref{e:avgv_G} are reasonable for the parameters used.

Note, that the pinning barrier which stops the domain wall motion can be overcome by thermal fluctuations. The role of these fluctuations was investigated by Duine et al. \cite{duinePRL07} within the framework of an extended, stochastic LLG equation.  These fluctuations are relevant in the limit of very thin wires where by thermal activation the pinning potential can be overcome on sufficiently long time scales, leading to a finite domain wall motion even below the critical current. In our work, however, fluctuations are not considered so that the results are relevant for thicker wires where thermal fluctuations of the domain wall profile can be neglected.

\subsection{Non-adiabatic spin torque effects}             
In the following, non-adiabatic spin torque is taken into account and its effect is discussed in more detail. For comparison with previous investigations,\cite{thiavilleEPL05,schiebackEPJB07} the non-adiabatic prefactor $\bG$ is assumed to be temperature independent and is investigated in relation to the temperature independent microscopic damping constant $\lambda$. Note, however, that in Ref.~\onlinecite{xiaoPRB06} non-local contributions, which  are strongly correlated to the domain wall width, are predicted, which are neglected for our wide walls here.

\begin{figure}
\includegraphics[width=8cm]{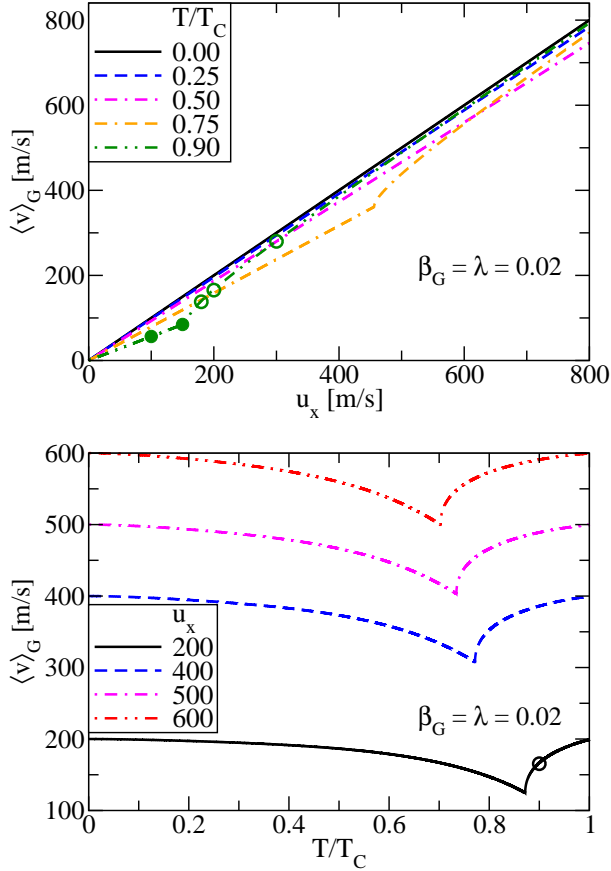}
\caption{(Color online) The top figure shows the average domain wall velocity $\langle v \rangle_{\rm G}$ calculated from the LLB equation
  with Gilbert damping vs. the effective spin current $u_x$ for different temperatures. The bottom figure shows $\langle v \rangle_{\rm G}$ vs. the reduced temperature $T/T_{\rm C}$. The lines represent the analytic solution (see Eq.~\ref{e:avgv_G}) and the points are from numerical simulations.  The calculations are for the non-adiabatic case, $\beta_{\rm G} = \lambda = 0.02$.}
\label{fig:dwVelocity_bG--0.020.eps}
\end{figure}

Our results for the case $\beta_{\rm G} = \lambda$ are shown in  Fig.~\ref{fig:dwVelocity_bG--0.020.eps}. In the zero temperature limit it is $m=1$ and $\alpha_{\perp} = \lambda$ so that the last term of Eq.~\ref{e:ex-LLG-LLB} vanishes and only the first spin torque term remains finite which is responsible for the displacement of the domain wall along the $x$-axis. The magnetization is, hence, not tilted out of the easy plane  and no torque occurs due to the precessional or relaxational part of the LLB equation. This behavior was already observed in previous numerical investigations of the LLG equation and is discussed in more detail in \onlinecite{thiavilleEPL05,schiebackEPJB07}.

In the case of elevated temperatures, the situation is different due to the fact that the last term of Eq.~\ref{e:ex-LLG-LLB} does not vanish because of the temperature dependence of $m$ and $\alpha_{\perp}$. This term is responsible for tilting the magnetization out of the easy plane and it leads to the existence of the Walker threshold $\uWG$ (Eq.~\ref{e:uWalker_G}).  Fig.~\ref{fig:dwVelocity_bG--0.020.eps} shows that  two regimes can be distinguished:  for $u_x < \uWG$ the velocity $\langle v \rangle_{\rm{G}}$ shows a linear behavior as in the zero temperature limit while in the regime $u_x > \uWG$ the second term in Eq.~\ref{e:avgv_G} takes over and the velocity increases even faster.  Here, the last term of Eq.~\ref{e:ex-LLG-LLB} leads to a continuous rotation of the magnetization around the $x$-axes. Following Eq.~\ref{e:uWalker_G} the transition between these regimes is shifted to lower effective spin currents with increasing temperature vanishing at \unit[0]{K}.

The averaged domain wall velocity as a function of temperature for different effective spin current values is shown in the bottom part of Fig.~\ref{fig:dwVelocity_bG--0.020.eps}.  First, it decreases with increasing temperature until a minimum value is reached, after which the velocity increases.  The minimum can be identified once again as the Walker threshold. This behavior is found for all effective velocities although the value of the minimum of the velocity is shifted to higher temperatures for lower effective velocities.

\begin{figure}
\includegraphics[width=8cm]{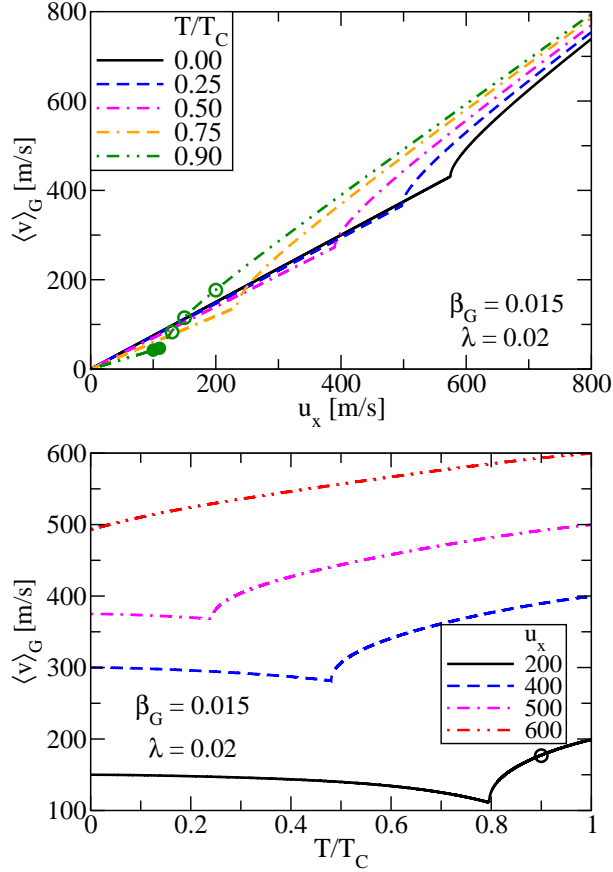}
\caption{(Color online) The top figure shows the average domain wall velocity $\langle v \rangle_{\rm G}$ calculated from the LLB equation
  with Gilbert damping vs. the effective spin current $u_x$ for different temperatures. The bottom figure shows $\langle v \rangle_{\rm G}$ vs. the reduced temperature $T/T_{\rm C}$. The lines represent the analytic solution (see Eq.~\ref{e:avgv_G}) and the points are from numerical simulations. The calculations are for the non-adiabatic case,  $\beta_{\rm G} < \lambda$.}
\label{fig:dwVelocity_bG--0.015.eps}
\end{figure}

Fig.~\ref{fig:dwVelocity_bG--0.015.eps} shows results for the case $\bG < \lambda$.  Here, the term responsible for tilting the magnetization out of the easy plane plays a crucial role for all temperatures even at \unit[0]{K}.  As before, the Walker threshold is shifted to lower effective velocities with increasing temperature.  In comparison to the case $\bG = \lambda$, the Walker threshold occurs at lower effective spin currents for the same temperature value so that the precession of the domain wall sets in earlier.

The averaged domain wall velocity as a function of the temperature is shown in the bottom part of Fig.~\ref{fig:dwVelocity_bG--0.015.eps} for different values of the effective spin current. As before a minimum exists which is shifted to lower temperatures for higher effective velocities, consistent with the shift of the Walker threshold discussed above. In comparison to  the $\bG = \lambda$  case, this shift of the minima is more pronounced.

\begin{figure}
\includegraphics[width=8cm]{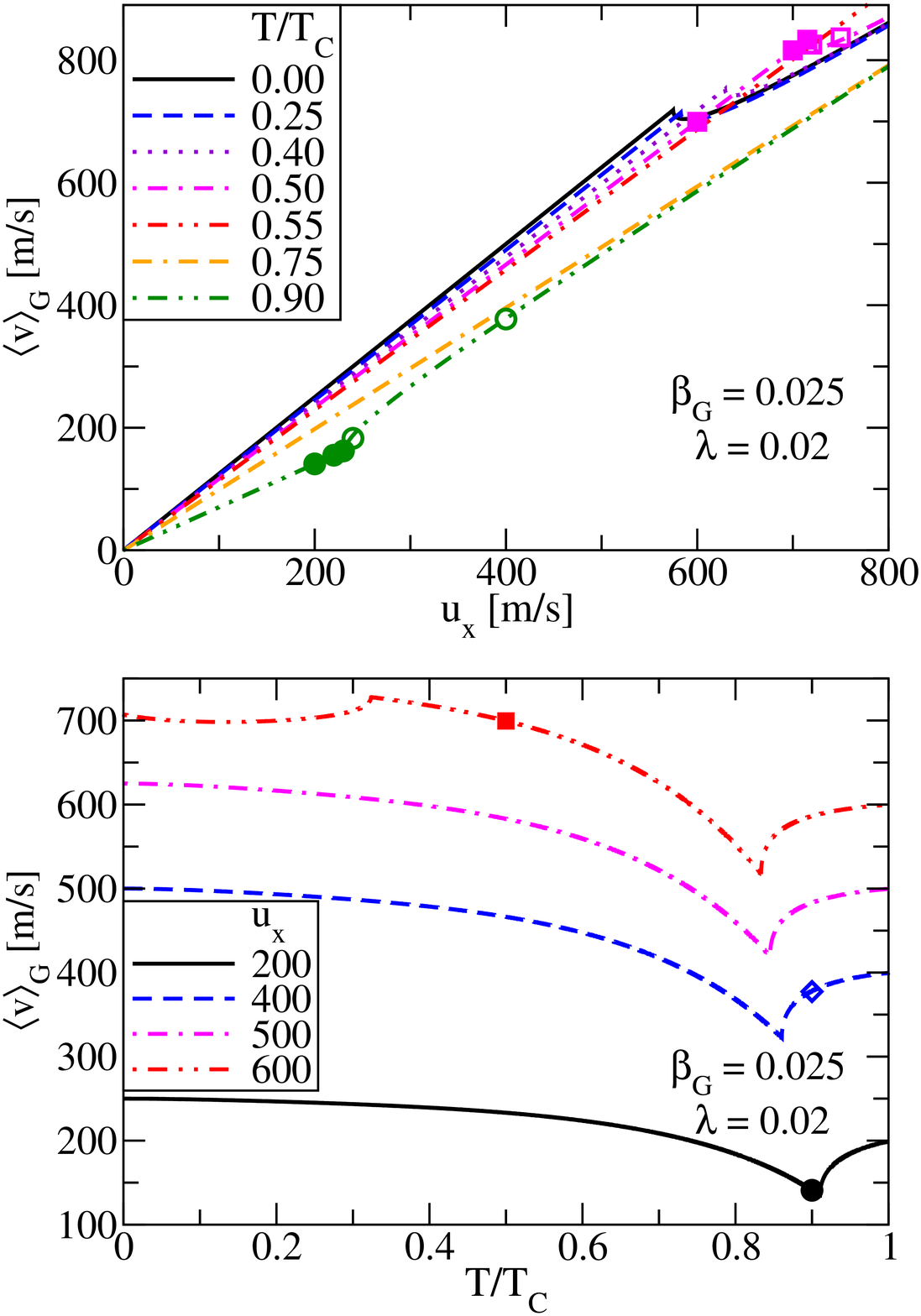}
\caption{(Color online) The top figure shows the average domain wall velocity $\langle v \rangle_{\rm G}$ calculated from the LLB equation
  with Gilbert damping vs. the effective spin current $u_x$ for different temperatures. The bottom figure shows $\langle v \rangle_{\rm G}$ vs. the reduced temperature $T/T_{\rm C}$. The lines represent the analytic solution (see Eq.~\ref{e:avgv_G}) and the points are from numerical simulations. For $\ux=\unit[600]{m/s}$ two crossings of the Walker threshold are visible ($T/\TC=0.32$ and $T/\TC=0.83$). The calculations are for the non-adiabatic case, $\beta_{\rm G} > \lambda$.}
\label{fig:dwVelocity_bG--0.025.eps}
\end{figure}

Finally, Fig.~\ref{fig:dwVelocity_bG--0.025.eps} shows results for the case $\bG > \lambda$. Here, the behavior differs from the two cases before. First of all, above the Walker threshold, the wall velocity increases slower than linear not faster as before. This is due to the sign change in Eq.~\ref{e:avgv_G}. Also, starting from low temperature the Walker threshold is first shifted to higher currents. In the effective spin current range shown in the figure the Walker threshold even disappears due to the fact that it is shifted  out of the range presented. Surprisingly, at temperatures close to $\TC$ another Walker threshold appears at lower effective velocities above which the averaged wall velocity increases faster than linear.

This second transition can also be identified from the averaged velocity as a function of temperature as shown in the bottom part of Fig.~\ref{fig:dwVelocity_bG--0.025.eps}.
For an effective spin current of $\unit[600]{m/s}$ there are two temperatures  where the behavior of the velocity changes. This corresponds to the fact that the Walker threshold as shown in Fig.~\ref{fig:fig_walker_breakdown} can be crossed twice for $\bG > \lambda$ and certain values of the effective current, leading to this intricate behavior, which could be easily identified if observed experimentally.

\subsection{Comparison of Gilbert and Landau-Lifshitz damping}     

\begin{figure}
\includegraphics[width=8cm]{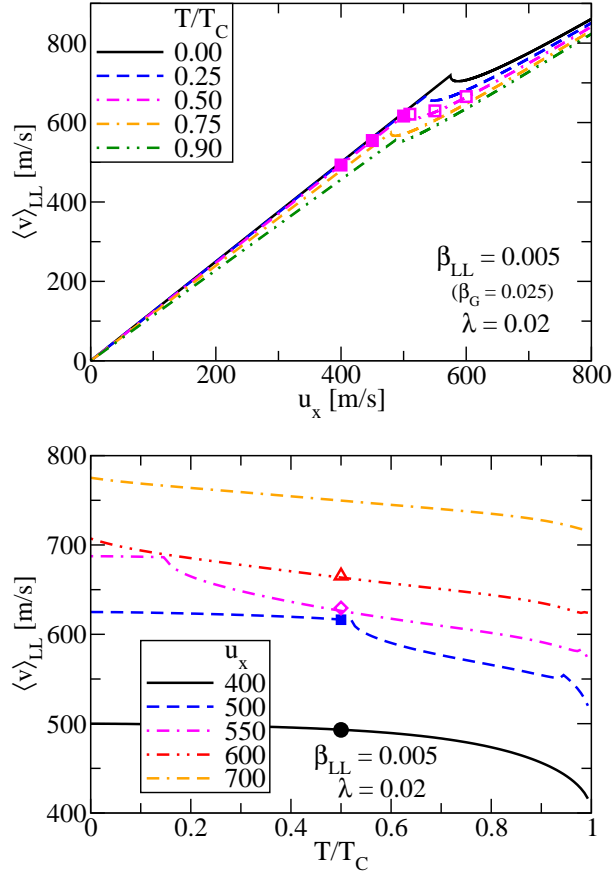}
\caption{(Color online) The top figure shows the average domain wall velocity $\langle v \rangle_{\rm G}$ calculated from the extended LLB equation with Landau-Lifshitz damping vs. the effective spin current $u_x$ for different temperatures. The bottom figure shows $\langle v \rangle_{\rm LL}$ vs. the reduced temperature $T/T_{\rm C}$. The lines represent analytic solutions and the points are from numerical simulations. The calculations are for the non-adiabatic case, non-adiabatic case $\beta_{\rm LL} = 0.005$ ($\beta_{\rm G} = 0.025)$ and $\lambda = 0.02$).}
\label{fig:dwVelocity_bLL--0.005.eps}
\end{figure}

In this subsection the difference between the assumption of Gilbert damping on the one hand and Landau-Lifshitz damping on the other hand is discussed for the case $\bG > \lambda$.  As mentioned before the prefactors for the non-adiabatic spin torque term can be transformed as $\bLL = \bG - \alpha_{\perp}/\me$ . This transformation is temperature dependent so that a qualitatively different temperature dependence exists for  Landau-Lifshitz damping in comparison to the Gilbert damping discussed before. Fig.~\ref{fig:dwVelocity_bLL--0.005.eps} summarizes our results for the extended LLB equation assuming Landau-Lifshitz damping (Eq.~\ref{e:LL-LLB}).

Comparing Fig.~\ref{fig:dwVelocity_bG--0.025.eps} and Fig.~\ref{fig:dwVelocity_bLL--0.005.eps} it is found that the curves for \unit[0]{K} are indeed the same.  However,  the behavior at elevated temperatures is completely different:  a Walker threshold can be identify for all temperatures. Below the threshold the domain wall is moved along the wire with an averaged velocity proportional to the effective spin current. Above the threshold additionally to this movement the magnetization precesses around the $x$-axis, so that the velocity of the domain wall is decreasing. The temperature dependent threshold is shifted to lower effective velocities for higher temperatures. Here, the \unit[0]{K} curve is an upper limit for the averaged domain wall velocity.

The behavior of the domain wall velocity as a function of the temperature is less complicated than in the corresponding  Gilbert case. For low currents the averaged velocity steadily decreases with increasing temperatures.   For slightly larger currents the domain wall moves with a constant velocity until the Walker threshold is reached. Above this point the velocity is decreasing. Close to the Curie temperature another transition point is reached where the domain wall velocity decreases even faster.  Here, the Walker threshold is crossed again (see Fig.~\ref{fig:fig_walker_breakdown}). A further increase of the effective spin current leads to a shift of the first transition point to lower temperatures and only two regimes can be identified.

\section{Conclusions}      
In summary, we combined the LL form of the LLB equation as well as the
Gilbert form of the LLB equation with the adiabatic and non-adiabatic
spin torque terms. We investigated analytically as well as numerically
domain wall motion at various temperatures for the adiabatic and
non-adiabatic cases. The Walker threshold as well as the domain wall
velocities show a strong temperature dependence. Furthermore, we found
a different behavior for the temperature dependent Walker threshold
assuming the Gilbert form of damping or the LL form. Since the two
behaviors can be qualitatively different, a measurement of the
temperature dependence of the velocity and the Walker threshold could
pose a unique opportunity to identify, whether the Gilbert or the LL
formulation are the physically relevant one. This then in turn could
have implications for determining the physically relevant description
of damping, which is one of the key open questions in magnetization
dynamics.

\begin{acknowledgments}
The authors thank N. Kazantseva for helpful discussions. This work was funded by the Deutsche Forschungsgemeinschaft (SFB 767), Landesstiftung Baden-W{\"u}rttemberg, and the ERC. Granting of
computer time from HLRS, and NIC is gratefully acknowledged.
\end{acknowledgments}


\begin{thebibliography}{31}
\expandafter\ifx\csname natexlab\endcsname\relax\def\natexlab#1{#1}\fi
\expandafter\ifx\csname bibnamefont\endcsname\relax
  \def\bibnamefont#1{#1}\fi
\expandafter\ifx\csname bibfnamefont\endcsname\relax
  \def\bibfnamefont#1{#1}\fi
\expandafter\ifx\csname citenamefont\endcsname\relax
  \def\citenamefont#1{#1}\fi
\expandafter\ifx\csname url\endcsname\relax
  \def\url#1{\texttt{#1}}\fi
\expandafter\ifx\csname urlprefix\endcsname\relax\def\urlprefix{URL }\fi
\providecommand{\bibinfo}[2]{#2}
\providecommand{\eprint}[2][]{\url{#2}}

\bibitem[{\citenamefont{Kl{\"a}ui}(2008)}]{klauiJP09}
\bibinfo{author}{\bibfnamefont{M.}~\bibnamefont{Kl{\"a}ui}},
  \bibinfo{journal}{J.~Phys.: Condens. Matter} \textbf{\bibinfo{volume}{20}},
  \bibinfo{pages}{313001} (\bibinfo{year}{2008}).

\bibitem[{\citenamefont{Yamaguchi et~al.}(2004)\citenamefont{Yamaguchi, Ono,
  Nasu, Miyake, Mibu, and Shinjo}}]{yamaguchiPRL04}
\bibinfo{author}{\bibfnamefont{A.}~\bibnamefont{Yamaguchi}},
  \bibinfo{author}{\bibfnamefont{T.}~\bibnamefont{Ono}},
  \bibinfo{author}{\bibfnamefont{S.}~\bibnamefont{Nasu}},
  \bibinfo{author}{\bibfnamefont{K.}~\bibnamefont{Miyake}},
  \bibinfo{author}{\bibfnamefont{K.}~\bibnamefont{Mibu}}, \bibnamefont{and}
  \bibinfo{author}{\bibfnamefont{T.}~\bibnamefont{Shinjo}},
  \bibinfo{journal}{Phys. Rev. Lett.} \textbf{\bibinfo{volume}{92}},
  \bibinfo{pages}{077205} (\bibinfo{year}{2004}).

\bibitem[{\citenamefont{Kl{\"{a}}ui et~al.}(2005)\citenamefont{Kl{\"{a}}ui,
  Vaz, Bland, Wernsdorfer, Faini, Cambril, Heyderman, Nolting, and
  R\"udiger}}]{klauiPRL05}
\bibinfo{author}{\bibfnamefont{M.}~\bibnamefont{Kl{\"{a}}ui}},
  \bibinfo{author}{\bibfnamefont{C.~A.~F.} \bibnamefont{Vaz}},
  \bibinfo{author}{\bibfnamefont{J.~A.~C.} \bibnamefont{Bland}},
  \bibinfo{author}{\bibfnamefont{W.}~\bibnamefont{Wernsdorfer}},
  \bibinfo{author}{\bibfnamefont{G.}~\bibnamefont{Faini}},
  \bibinfo{author}{\bibfnamefont{E.}~\bibnamefont{Cambril}},
  \bibinfo{author}{\bibfnamefont{L.~J.} \bibnamefont{Heyderman}},
  \bibinfo{author}{\bibfnamefont{F.}~\bibnamefont{Nolting}}, \bibnamefont{and}
  \bibinfo{author}{\bibfnamefont{U.}~\bibnamefont{R\"udiger}},
  \bibinfo{journal}{Phys. Rev. Lett.} \textbf{\bibinfo{volume}{94}},
  \bibinfo{pages}{106601} (\bibinfo{year}{2005}).

\bibitem[{\citenamefont{Heyne et~al.}(2008)\citenamefont{Heyne, Kl{\"a}ui,
  Backes, Moore, Krzyk, R{\"u}diger, Heyderman, Rodr{\'{\i}}guez, Nolting,
  Mentes et~al.}}]{heynePRL08}
\bibinfo{author}{\bibfnamefont{L.}~\bibnamefont{Heyne}},
  \bibinfo{author}{\bibfnamefont{M.}~\bibnamefont{Kl{\"a}ui}},
  \bibinfo{author}{\bibfnamefont{D.}~\bibnamefont{Backes}},
  \bibinfo{author}{\bibfnamefont{T.~A.} \bibnamefont{Moore}},
  \bibinfo{author}{\bibfnamefont{S.}~\bibnamefont{Krzyk}},
  \bibinfo{author}{\bibfnamefont{U.}~\bibnamefont{R{\"u}diger}},
  \bibinfo{author}{\bibfnamefont{L.~J.} \bibnamefont{Heyderman}},
  \bibinfo{author}{\bibfnamefont{A.~F.} \bibnamefont{Rodr{\'{\i}}guez}},
  \bibinfo{author}{\bibfnamefont{F.}~\bibnamefont{Nolting}},
  \bibinfo{author}{\bibfnamefont{T.~O.} \bibnamefont{Mentes}},
  \bibnamefont{et~al.}, \bibinfo{journal}{Phys. Rev. Lett.}
  \textbf{\bibinfo{volume}{100}}, \bibinfo{pages}{066603}
  (\bibinfo{year}{2008}).

\bibitem[{\citenamefont{Li and Zhang}(2004{\natexlab{a}})}]{liPRL04}
\bibinfo{author}{\bibfnamefont{Z.}~\bibnamefont{Li}} \bibnamefont{and}
  \bibinfo{author}{\bibfnamefont{S.}~\bibnamefont{Zhang}},
  \bibinfo{journal}{Phys. Rev. Lett.} \textbf{\bibinfo{volume}{92}},
  \bibinfo{pages}{207203} (\bibinfo{year}{2004}{\natexlab{a}}).

\bibitem[{\citenamefont{Thiaville et~al.}(2005)\citenamefont{Thiaville,
  Nakatani, Miltat, and Suzuki}}]{thiavilleEPL05}
\bibinfo{author}{\bibfnamefont{A.}~\bibnamefont{Thiaville}},
  \bibinfo{author}{\bibfnamefont{Y.}~\bibnamefont{Nakatani}},
  \bibinfo{author}{\bibfnamefont{J.}~\bibnamefont{Miltat}}, \bibnamefont{and}
  \bibinfo{author}{\bibfnamefont{N.}~\bibnamefont{Suzuki}},
  \bibinfo{journal}{Europhys. Lett.} \textbf{\bibinfo{volume}{69}},
  \bibinfo{pages}{990} (\bibinfo{year}{2005}).

\bibitem[{\citenamefont{Laufenberg et~al.}(2006)\citenamefont{Laufenberg,
  B\"uhrer, Bedau, Melchy, Kl\"aui, Vila, Faini, Vaz, Bland, and
  R\"udiger}}]{laufenbergPRL06}
\bibinfo{author}{\bibfnamefont{M.}~\bibnamefont{Laufenberg}},
  \bibinfo{author}{\bibfnamefont{W.}~\bibnamefont{B\"uhrer}},
  \bibinfo{author}{\bibfnamefont{D.}~\bibnamefont{Bedau}},
  \bibinfo{author}{\bibfnamefont{P.-E.} \bibnamefont{Melchy}},
  \bibinfo{author}{\bibfnamefont{M.}~\bibnamefont{Kl\"aui}},
  \bibinfo{author}{\bibfnamefont{L.}~\bibnamefont{Vila}},
  \bibinfo{author}{\bibfnamefont{G.}~\bibnamefont{Faini}},
  \bibinfo{author}{\bibfnamefont{C.~A.~F.} \bibnamefont{Vaz}},
  \bibinfo{author}{\bibfnamefont{J.~A.~C.} \bibnamefont{Bland}},
  \bibnamefont{and}
  \bibinfo{author}{\bibfnamefont{U.}~\bibnamefont{R\"udiger}},
  \bibinfo{journal}{Phys. Rev. Lett.} \textbf{\bibinfo{volume}{97}},
  \bibinfo{pages}{046602} (\bibinfo{year}{2006}).

\bibitem[{\citenamefont{Berger}(1978)}]{bergerJAP78}
\bibinfo{author}{\bibfnamefont{L.}~\bibnamefont{Berger}}, \bibinfo{journal}{J.
  Appl. Phys.} \textbf{\bibinfo{volume}{49}}, \bibinfo{pages}{2156}
  (\bibinfo{year}{1978}).

\bibitem[{\citenamefont{Slonczewski}(1996)}]{slonczewskiJMMM96}
\bibinfo{author}{\bibfnamefont{J.~C.} \bibnamefont{Slonczewski}},
  \bibinfo{journal}{J. Magn. Magn. Mat.} \textbf{\bibinfo{volume}{159}},
  \bibinfo{pages}{L1} (\bibinfo{year}{1996}).

\bibitem[{\citenamefont{{D.\,A.~Garanin}}(1997)}]{garaninPRB97}
\bibinfo{author}{\bibnamefont{{D.\,A.~Garanin}}}, \bibinfo{journal}{Phys. Rev.
  B} \textbf{\bibinfo{volume}{55}}, \bibinfo{pages}{3050}
  (\bibinfo{year}{1997}).

\bibitem[{\citenamefont{Kazantseva et~al.}(2008)\citenamefont{Kazantseva,
  Hinzke, Nowak, Chantrell, Atxitia, and Chubykalo-Fesenko}}]{kazantsevaPRB08}
\bibinfo{author}{\bibfnamefont{N.}~\bibnamefont{Kazantseva}},
  \bibinfo{author}{\bibfnamefont{D.}~\bibnamefont{Hinzke}},
  \bibinfo{author}{\bibfnamefont{U.}~\bibnamefont{Nowak}},
  \bibinfo{author}{\bibfnamefont{R.~W.} \bibnamefont{Chantrell}},
  \bibinfo{author}{\bibfnamefont{U.}~\bibnamefont{Atxitia}}, \bibnamefont{and}
  \bibinfo{author}{\bibfnamefont{O.}~\bibnamefont{Chubykalo-Fesenko}},
  \bibinfo{journal}{Phys. Rev. B} \textbf{\bibinfo{volume}{77}},
  \bibinfo{pages}{184428} (\bibinfo{year}{2008}).

\bibitem[{\citenamefont{K{\"o}tzler et~al.}(1993)\citenamefont{K{\"o}tzler,
  Garanin, Hartl, and Jahn}}]{koetzlerPRL93}
\bibinfo{author}{\bibfnamefont{J.}~\bibnamefont{K{\"o}tzler}},
  \bibinfo{author}{\bibfnamefont{D.~A.} \bibnamefont{Garanin}},
  \bibinfo{author}{\bibfnamefont{M.}~\bibnamefont{Hartl}}, \bibnamefont{and}
  \bibinfo{author}{\bibfnamefont{L.}~\bibnamefont{Jahn}},
  \bibinfo{journal}{Phys. Rev. Lett.} \textbf{\bibinfo{volume}{71}},
  \bibinfo{pages}{177} (\bibinfo{year}{1993}).

\bibitem[{\citenamefont{Kazantseva et~al.}(2005)\citenamefont{Kazantseva,
  Wieser, and Nowak}}]{kazantsevaPRL05}
\bibinfo{author}{\bibfnamefont{N.}~\bibnamefont{Kazantseva}},
  \bibinfo{author}{\bibfnamefont{R.}~\bibnamefont{Wieser}}, \bibnamefont{and}
  \bibinfo{author}{\bibfnamefont{U.}~\bibnamefont{Nowak}},
  \bibinfo{journal}{Phys. Rev. Lett.} \textbf{\bibinfo{volume}{94}},
  \bibinfo{pages}{037206} (\bibinfo{year}{2005}).

\bibitem[{\citenamefont{Hinzke et~al.}(2008)\citenamefont{Hinzke, Kazantseva,
  Nowak, Mryasov, Asselin, and Chantrell}}]{hinzkePRB08}
\bibinfo{author}{\bibfnamefont{D.}~\bibnamefont{Hinzke}},
  \bibinfo{author}{\bibfnamefont{N.}~\bibnamefont{Kazantseva}},
  \bibinfo{author}{\bibfnamefont{U.}~\bibnamefont{Nowak}},
  \bibinfo{author}{\bibfnamefont{O.~N.} \bibnamefont{Mryasov}},
  \bibinfo{author}{\bibfnamefont{P.}~\bibnamefont{Asselin}}, \bibnamefont{and}
  \bibinfo{author}{\bibfnamefont{R.~W.} \bibnamefont{Chantrell}},
  \bibinfo{journal}{Phys. Rev. B} \textbf{\bibinfo{volume}{77}},
  \bibinfo{pages}{094407} (\bibinfo{year}{2008}).

\bibitem[{\citenamefont{Chubykalo-Fesenko
  et~al.}(2006)\citenamefont{Chubykalo-Fesenko, Nowak, Chantrell, and
  Garanin}}]{chubykaloPRB06}
\bibinfo{author}{\bibfnamefont{O.}~\bibnamefont{Chubykalo-Fesenko}},
  \bibinfo{author}{\bibfnamefont{U.}~\bibnamefont{Nowak}},
  \bibinfo{author}{\bibfnamefont{R.~W.} \bibnamefont{Chantrell}},
  \bibnamefont{and} \bibinfo{author}{\bibfnamefont{D.}~\bibnamefont{Garanin}},
  \bibinfo{journal}{Phys. Rev. B} \textbf{\bibinfo{volume}{74}},
  \bibinfo{pages}{094436} (\bibinfo{year}{2006}).

\bibitem[{\citenamefont{Kazantseva et~al.}(2009)\citenamefont{Kazantseva,
  Hinzke, Chantrell, and Nowak}}]{kazantsevaEPL09}
\bibinfo{author}{\bibfnamefont{N.}~\bibnamefont{Kazantseva}},
  \bibinfo{author}{\bibfnamefont{D.}~\bibnamefont{Hinzke}},
  \bibinfo{author}{\bibfnamefont{R.~W.} \bibnamefont{Chantrell}},
  \bibnamefont{and} \bibinfo{author}{\bibfnamefont{U.}~\bibnamefont{Nowak}},
  \bibinfo{journal}{Europhys. Lett.} \textbf{\bibinfo{volume}{86}},
  \bibinfo{pages}{27006} (\bibinfo{year}{2009}).

\bibitem[{\citenamefont{Bazaliy et~al.}(1998)\citenamefont{Bazaliy, Jones, and
  Zhang}}]{bazaliyPRB98}
\bibinfo{author}{\bibfnamefont{Y.~B.} \bibnamefont{Bazaliy}},
  \bibinfo{author}{\bibfnamefont{B.~A.} \bibnamefont{Jones}}, \bibnamefont{and}
  \bibinfo{author}{\bibfnamefont{S.~C.} \bibnamefont{Zhang}},
  \bibinfo{journal}{Phys. Rev. B} \textbf{\bibinfo{volume}{57}},
  \bibinfo{pages}{R3213} (\bibinfo{year}{1998}).

\bibitem[{\citenamefont{Thiaville et~al.}(2004)\citenamefont{Thiaville,
  Nakatani, Miltat, and Vernier}}]{thiavilleJAP04}
\bibinfo{author}{\bibfnamefont{A.}~\bibnamefont{Thiaville}},
  \bibinfo{author}{\bibfnamefont{Y.}~\bibnamefont{Nakatani}},
  \bibinfo{author}{\bibfnamefont{J.}~\bibnamefont{Miltat}}, \bibnamefont{and}
  \bibinfo{author}{\bibfnamefont{N.}~\bibnamefont{Vernier}},
  \bibinfo{journal}{J. Appl. Phys.} \textbf{\bibinfo{volume}{95}},
  \bibinfo{pages}{7049} (\bibinfo{year}{2004}).

\bibitem[{\citenamefont{Li and Zhang}(2004{\natexlab{b}})}]{liPRB04}
\bibinfo{author}{\bibfnamefont{Z.}~\bibnamefont{Li}} \bibnamefont{and}
  \bibinfo{author}{\bibfnamefont{S.}~\bibnamefont{Zhang}},
  \bibinfo{journal}{Phys. Rev. B} \textbf{\bibinfo{volume}{70}},
  \bibinfo{pages}{024417} (\bibinfo{year}{2004}{\natexlab{b}}).

\bibitem[{\citenamefont{He et~al.}(2006)\citenamefont{He, Li, and
  Zhang}}]{hePRB06}
\bibinfo{author}{\bibfnamefont{J.}~\bibnamefont{He}},
  \bibinfo{author}{\bibfnamefont{Z.}~\bibnamefont{Li}}, \bibnamefont{and}
  \bibinfo{author}{\bibfnamefont{S.}~\bibnamefont{Zhang}},
  \bibinfo{journal}{Phys. Rev. B} \textbf{\bibinfo{volume}{73}},
  \bibinfo{pages}{184408} (\bibinfo{year}{2006}).

\bibitem[{\citenamefont{Haney and Stiles}(2009)}]{haneyARXIV09}
\bibinfo{author}{\bibfnamefont{P.~M.} \bibnamefont{Haney}} \bibnamefont{and}
  \bibinfo{author}{\bibfnamefont{M.~D.} \bibnamefont{Stiles}}
  (\bibinfo{year}{2009}), \bibinfo{note}{http://arxiv.org/pdf/0906.2423}.

\bibitem[{\citenamefont{Stiles et~al.}(2007)\citenamefont{Stiles, Saslow,
  Donahue, and Zangwill}}]{stilesPRB07}
\bibinfo{author}{\bibfnamefont{M.~D.} \bibnamefont{Stiles}},
  \bibinfo{author}{\bibfnamefont{W.~M.} \bibnamefont{Saslow}},
  \bibinfo{author}{\bibfnamefont{M.~J.} \bibnamefont{Donahue}},
  \bibnamefont{and} \bibinfo{author}{\bibfnamefont{A.}~\bibnamefont{Zangwill}},
  \bibinfo{journal}{Phys. Rev. B} \textbf{\bibinfo{volume}{75}},
  \bibinfo{pages}{214423} (\bibinfo{year}{2007}).

\bibitem[{\citenamefont{Smith}(2008)}]{smithPRB08}
\bibinfo{author}{\bibfnamefont{N.}~\bibnamefont{Smith}},
  \bibinfo{journal}{Phys. Rev. B} \textbf{\bibinfo{volume}{78}},
  \bibinfo{pages}{216401} (\bibinfo{year}{2008}).

\bibitem[{\citenamefont{Stiles et~al.}(2008)\citenamefont{Stiles, Saslow,
  Donahue, and Zangwill}}]{stilesPRB08}
\bibinfo{author}{\bibfnamefont{M.~D.} \bibnamefont{Stiles}},
  \bibinfo{author}{\bibfnamefont{W.~M.} \bibnamefont{Saslow}},
  \bibinfo{author}{\bibfnamefont{M.~J.} \bibnamefont{Donahue}},
  \bibnamefont{and} \bibinfo{author}{\bibfnamefont{A.}~\bibnamefont{Zangwill}},
  \bibinfo{journal}{Phys. Rev. B} \textbf{\bibinfo{volume}{78}},
  \bibinfo{pages}{216402} (\bibinfo{year}{2008}).

\bibitem[{\citenamefont{{N.\,L.~Schryer} and
  {L.\,R.~Walker}}(1974)}]{schryerJAP74}
\bibinfo{author}{\bibnamefont{{N.\,L.~Schryer}}} \bibnamefont{and}
  \bibinfo{author}{\bibnamefont{{L.\,R.~Walker}}}, \bibinfo{journal}{J. Appl.
  Phys.} \textbf{\bibinfo{volume}{45}}, \bibinfo{pages}{5406}
  (\bibinfo{year}{1974}).

\bibitem[{\citenamefont{Garc\'{\i}a-Palacios and L\'azaro}(1998)}]{garciaPRB98}
\bibinfo{author}{\bibfnamefont{J.~L.} \bibnamefont{Garc\'{\i}a-Palacios}}
  \bibnamefont{and} \bibinfo{author}{\bibfnamefont{F.~J.}
  \bibnamefont{L\'azaro}}, \bibinfo{journal}{Phys. Rev. B}
  \textbf{\bibinfo{volume}{58}}, \bibinfo{pages}{14937} (\bibinfo{year}{1998}).

\bibitem[{\citenamefont{Nowak}(2001)}]{nowakARCP01}
\bibinfo{author}{\bibfnamefont{U.}~\bibnamefont{Nowak}}, in
  \emph{\bibinfo{booktitle}{Annual Reviews of Computational Physics IX}},
  edited by \bibinfo{editor}{\bibfnamefont{D.}~\bibnamefont{Stauffer}}
  (\bibinfo{publisher}{World Scientific}, \bibinfo{address}{Singapore},
  \bibinfo{year}{2001}), p. \bibinfo{pages}{105}.

\bibitem[{\citenamefont{Tatara and Kohno}(2004)}]{tataraPRL04}
\bibinfo{author}{\bibfnamefont{G.}~\bibnamefont{Tatara}} \bibnamefont{and}
  \bibinfo{author}{\bibfnamefont{H.}~\bibnamefont{Kohno}},
  \bibinfo{journal}{Phys. Rev. Lett.} \textbf{\bibinfo{volume}{92}},
  \bibinfo{pages}{086601} (\bibinfo{year}{2004}).

\bibitem[{\citenamefont{Schieback et~al.}(2007)\citenamefont{Schieback,
  Kl{\"a}ui, Nowak, R{\"u}diger, and Nielaba}}]{schiebackEPJB07}
\bibinfo{author}{\bibfnamefont{C.}~\bibnamefont{Schieback}},
  \bibinfo{author}{\bibfnamefont{M.}~\bibnamefont{Kl{\"a}ui}},
  \bibinfo{author}{\bibfnamefont{U.}~\bibnamefont{Nowak}},
  \bibinfo{author}{\bibfnamefont{U.}~\bibnamefont{R{\"u}diger}},
  \bibnamefont{and} \bibinfo{author}{\bibfnamefont{P.}~\bibnamefont{Nielaba}},
  \bibinfo{journal}{Euro. Phys. J. B} \textbf{\bibinfo{volume}{59}},
  \bibinfo{pages}{429} (\bibinfo{year}{2007}).

\bibitem[{\citenamefont{Duine et~al.}(2007)\citenamefont{Duine, N{\'u}{\~n}ez,
  and MacDonald}}]{duinePRL07}
\bibinfo{author}{\bibfnamefont{R.~A.} \bibnamefont{Duine}},
  \bibinfo{author}{\bibfnamefont{A.~S.} \bibnamefont{N{\'u}{\~n}ez}},
  \bibnamefont{and} \bibinfo{author}{\bibfnamefont{A.~H.}
  \bibnamefont{MacDonald}}, \bibinfo{journal}{Phys. Rev. Lett.}
  \textbf{\bibinfo{volume}{98}}, \bibinfo{pages}{056605}
  (\bibinfo{year}{2007}).

\bibitem[{\citenamefont{Xiao et~al.}(2006)\citenamefont{Xiao, Zangwill, and
  Stiles}}]{xiaoPRB06}
\bibinfo{author}{\bibfnamefont{J.}~\bibnamefont{Xiao}},
  \bibinfo{author}{\bibfnamefont{A.}~\bibnamefont{Zangwill}}, \bibnamefont{and}
  \bibinfo{author}{\bibfnamefont{M.~D.} \bibnamefont{Stiles}},
  \bibinfo{journal}{Phys. Rev. B} \textbf{\bibinfo{volume}{73}},
  \bibinfo{pages}{054428} (\bibinfo{year}{2006}).

\end{thebibliography}
\end{document}